\RequirePackage{ifpdf}
\ifpdf 
\documentclass[pdftex]{sigma}
\else
\documentclass{sigma}
\fi

\begin{document}
\allowdisplaybreaks

\renewcommand{\PaperNumber}{010}

\FirstPageHeading

\ShortArticleName{Full Kostant--Toda Lattice}

\ArticleName{A Gentle (without Chopping) Approach\\ to the Full Kostant--Toda Lattice}

\Author{Pantelis A.  DAMIANOU~$^\dag$ and Franco MAGRI~$^\ddag$}
\AuthorNameForHeading{P.A. Damianou and F. Magri}

\Address{$^\dag$~Department of Mathematics and Statistics, University 
of Cyprus, 1678, Nicosia, Cyprus} 
\EmailD{\href{mailto:damianou@ucy.ac.cy}{damianou@ucy.ac.cy}}
\URLaddressD{\href{http://www.ucy.ac.cy/~damianou/}{http://www.ucy.ac.cy/\~{}damianou/}} 
\Address{$^\ddag$~Department of Mathematics, University of Milano Bicocca, Via Corsi 58,\\
$\phantom{^\ddag}$~I 20126 Milano, Italy}
\EmailD{\href{mailto:magri@matapp.unimib.it}{magri@matapp.unimib.it}}

\ArticleDates{Received September 22, 2005, in final form October 24, 2005; Published online October 25, 2005}

\Abstract{In this paper we propose a new algorithm for obtaining
the rational integrals of the full Kostant--Toda lattice. This new
approach is based on a reduction of a bi-Hamiltonian system on
$gl(n, {\mathbb R})$. This system was obtained by reducing the space
of maps from $Z_n$ to $GL(n, {\mathbb R})$ endowed with a structure of
a pair of Lie-algebroids.}


\Classification{37J35; 70H06} 

\newcommand {\ds} {\displaystyle \sum}
\def\p{ \partial}

\section{Introduction}
The Toda lattice is arguably the most fundamental and basic of all
finite dimensional integrable systems. It has various intriguing
connections with other parts of mathematics and physics.

 The Hamiltonian of the Toda lattice is given by 
\begin{gather}
  H(q_1, \dots, q_N,   p_1, \dots, p_N) 
  = \sum_{i=1}^N   { 1 \over 2} \, p_i^2 +
\sum _{i=1}^{N-1}  e^{ q_i-q_{i+1}} . \label{a1} \end{gather}

Equation (\ref{a1}) is   known as  the classical, finite,
non-periodic Toda lattice to distinguish the system from the many
and  various other versions, e.g.,  the relativistic, quantum,
periodic etc.  This system was investigated in
  \cite{flaschka1,flaschka2,henon,manakov,moser,moser2,toda}
and numerous of other papers; see~\cite{damianou4} for a more
extensive bibliography.

Hamilton's equations become
\begin{gather*}
\dot q_j =p_j,    \\
\dot p_j=e^{ q_{j-1}-q_j }- e^{q_j- q_{j+1}}.
\end{gather*}
The system is  integrable. One can find a set of
independent functions $\{  H_1, \dots,  H_N \} $  which are
constants of motion for Hamilton's equations. To  determine the
constants of motion, one uses Flaschka's transformation: 
\begin{gather}
  a_i  = {1 \over 2} e^{ {1 \over 2} (q_i - q_{i+1} ) } , \qquad
             b_i  = -{ 1 \over 2} p_i .     \label{a2}
\end{gather}
Then
\begin{gather}
 \dot a _i =  a_i  (b_{i+1} -b_i ),\qquad  \dot b _i =  2  ( a_i^2 - a_{i-1}^2 ) .  \label{a3}
\end{gather}
 These equations can be written as a Lax pair  $\dot L =
[B, L] $, where $L$ is the Jacobi matrix
\begin{gather}
 L= \begin{pmatrix}  b_1 &  a_1 & 0 & \cdots & \cdots & 0 \cr
                   a_1 & b_2 & a_2 & \cdots &    & \vdots \cr
                   0 & a_2 & b_3 & \ddots &  &  \cr
                   \vdots & & \ddots & \ddots & & \vdots \cr
                   \vdots & & & \ddots & \ddots & a_{N-1} \cr
                   0 & \cdots & & \cdots & a_{N-1} & b_N   \cr
                   \end{pmatrix}
                   ,\label{a4}
\end{gather}

\smallskip
\noindent and $B$ is the skew-symmetric part of $L$.  This is an
example of an isospectral deformation; the entries of $L$ vary
over time but the eigenvalues  remain constant.
 It follows that the  functions $ H_i={1 \over i} \,{\rm tr} \, L^i$ are  constants of
motion.

Note that the Lax pair (\ref{a4}) has the form  
\[ 
\dot L(t)=[P\,
L(t), L(t)],
 \]
where $P$ denotes  the projection onto the skew-symmetric part  in
the decomposition of $L$ into skew-symmetric plus lower
triangular.

The Toda lattice was generalized in several directions.

We mention the Bogoyavlensky--Toda lattices which  generalize the
 Toda lattice (which corresponds to a root system of type $A_n$)
to other simple Lie groups. This generalization is due to
Bogoyavlensky~\cite{bogo}. These systems were studied extensively
in~\cite{kostant} where the solution of the systems  was connected intimately with the representation
theory of simple Lie groups.

Another generalization  is due to    Deift, Li, Nanda and
 Tomei \cite{deift} who showed that the system remains integrable when $L$ is replaced
 by a full (generic) symmetric $n \times n$ matrix.

 Another variation is the
full Kostant--Toda lattice (FKT) \cite{ercolani,flaschka3,singer}. We briefly describe the system:
   In~\cite{kostant} Kostant conjugates the matrix $L$ in (\ref{a4})  by a diagonal matrix
  to obtain a matrix of the form
\begin{gather}
L= \begin{pmatrix} b_1 &  1 & 0 & \cdots & \cdots & 0 \cr
                   a_1 & b_2 & 1 & \ddots &    & \vdots \cr
                   0 & a_2 & b_3 & \ddots &  &  \vdots \cr
                   \vdots & \ddots & \ddots & \ddots & & 0 \cr
                   \vdots & & & \ddots & \ddots & 1 \cr
                   0 & \cdots &  \cdots & 0 & a_{n-1} & b_n   \cr \end{pmatrix} .
\end{gather}

The equations take the form
\[ 
\dot X(t)=[ X(t), P  X(t)],
 \] where $P$ is the projection
onto the strictly lower triangular part of $X(t)$. This form is
convenient in applying Lie theoretic techniques to describe the
system.

To obtain the full Kostant--Toda lattice we fill the lower
triangular
 part of $L$ in \eqref{a4} with  additional variables. ($P $ is again the projection
 onto the strictly lower part of $X(t)$).  So, using the notation from \cite{ercolani,flaschka3} and
 \cite{singer}
\[ 
\dot X(t)=[ X(t), P 
 X(t)], 
 \] where $X$ is in $
\epsilon + B_-$ and $ P  X$ is in $N_-$. $B_-$ is the Lie
algebra of lower triangular matrices and~$N_-$ is the Lie algebra
of strictly lower triangular matrices.  The fixed matrix
$\epsilon$ has a general form in terms of root systems:
\[ 
\epsilon = \sum_{\alpha \in \Delta} x_{\alpha}, 
\] where
$\Delta $ denotes the set of  simple roots. In fact the FKT
lattice itself can easily be generalized for each simple Lie
group; see  \cite{kyriakos,ercolani}.

 In the case of $sl(4,{\mathbb C})$  the matrix $X$
 has   the form
\begin{equation}
X= \begin{pmatrix} f_1 &  1 & 0 & 0 \cr
                   g_1 & f_2 & 1 &  0 \cr
                   h_1 & g_2 & f_3 & 1   \cr
                   k_1 & h_2 & g_3 & f_4   \cr \end{pmatrix} ,
\end{equation}
with $\sum\limits_i f_i =0$.

 The functions  $H_i= { 1 \over i} \, {\rm Tr} \ X^i$  are still in involution but they are  not enough to
ensure integrability.  This is a crucial point: the existence of a
 Lax pair does not guarantee integrability. There are, however, additional integrals which are rational
functions of the entries of $X$. The method used to obtain these
additional integrals is called chopping and was used originally 
in~\cite{deift}  for the full symmetric Toda and  later 
in~\cite{ercolani} for the case of the full Kostant--Toda lattice.

 In this paper we
use a different method of obtaining these rational integrals which
does not involve chopping.  This method uses a reduction of a
bi--Hamiltonian system on $gl(N, {\mathbb R})$,  a system which was
first defined  in \cite{meucci1,meucci2}.  In~\cite{meucci2}
 A.~Meucci presents the bi-Hamiltonian structure of Toda$_3$, a
 dynamical system studied by Kupershmidt  in~\cite{kup}  as a reduction of
 the KP hierarchy. Meucci  derives this structure by a suitable
 restriction of the set of maps from $Z_d$, where $Z_d$ is the cyclic group of order $d$,
  to $GL(3, {\mathbb R})$, in
 the context of Lie algebroids.

 In \cite{meucci1}  the bi-Hamiltonian structure of the periodic Toda lattice
 is investigated by the reduction process described above using
 maps from  $Z_d$ to $GL(2, {\mathbb R})$.  This approach parallels the
 work of~\cite{falqui}  where the continuous analog of the Toda lattice is
 studied, namely the KdV.  If instead the target space is  $gl(3,
 {\mathbb R})$ one obtains the Boussineq hierarchy.  The work of
 Meucci is a discrete version of this approach.  If one
 generalizes the cases $N=2,3$, i.e.\ consider maps from $Z_d$ to
 $GL(N, {\mathbb R})$ the resulting system will be denoted by Toda$_N$.
 In the present paper we use the results of \cite{meucci1,meucci2}  as a starting
 point. We begin with the bi-Hamiltonian system obtained on
 $gl(N, {\mathbb R})$ in the particular case $d=N$  and use a further reduction to obtain the
 dynamics of the full Kostant--Toda lattice.  We then propose a
 new algorithm which produces all the rational integrals for the
 FKT lattice without using chopping. We present
 this algorithm by specific examples ($N=3,4$) in Section 4.
 Sections 2 and 3 contain  a general review of the old methods and results
 on integrability and bi-Hamiltonian structure  of the FKT lattice.

\section{Integrability of the FKT lattice}

  Let ${\cal G}= sl(n)$, the Lie algebra of $n \times n$ matrices of trace zero.
 Using the decomposition  ${\cal G}= B_+  \oplus  N_-$  we can identify   $B_+^*$ with the
annihilator of $N_-$  with respect to the Killing  form. This
annihilator is $B_-$.  Thus we can identify $B_+^*$ with $B_-$ and
therefore with $\epsilon +B_-$ as well.

 The Lie--Poisson bracket in
 the case of $sl(4, {\mathbb C} )$ is given by the  following defining relations:
\begin{gather*}
\{ g_i, g_{i+1} \} = h_i,\\
\{g_i, f_i \} = - g_i,\\
\{ g_i, f_{i+1} \}= g_i,\\
\{ h_i, f_i \} =-h_i,\\
\{  h_i, f_{i+2} \}= h_i,\\
\{ g_1, h_2 \} = k_1,\\
\{ g_3, h_1 \} =- k_1,\\
\{ k_1, f_1 \} = -k_1,\\
\{ k_1, f_4 \} =k_1.
\end{gather*}
All other brackets are zero. Actually,  we calculated
the brackets on $gl(4, {\mathbb C})$; the trace of $X$ now becomes a
Casimir. The Hamiltonian in this bracket is $H_2 = { 1 \over 2}\,
{ \rm Tr}\,  X^2$.

\begin{remark} If we use a more conventional notation for
the matrix $X$, i.e.\ $x_{ij}$ for  $i \ge j$, $x_{ii+1}=1$, and
all other entries zero, then the bracket is simply 
\begin{gather} 
\{ x_{ij}, x_{kl} \} = \delta_{li} x_{kj} -\delta_{jk} x_{il} . \label{a5}
\end{gather}
\end{remark}

 The functions  $H_i= { 1 \over i} \, {\rm Tr} \, X^i$  are still in involution but they are  not enough to
ensure integrability. There are, however,  additional integrals
and the interesting feature of  this  system is that the additional integrals turn out to be rational
functions of the entries of $X$.  We describe the constants of
motion following references~\cite{ercolani,flaschka3,singer}.

 For $k=0, \dots , \left[ { (n-1) \over 2}\right]$, denote by $( X-
\lambda \, { \rm Id})_{ (k)}$ the result of removing the first $k$
rows and last $k$ columns from $X- \lambda \,{\rm  Id}$, and let
\[ \det
 ( X- \lambda \, { \rm Id})_{ (k)} = E_{0k} \lambda
^{n- 2k} + \dots + E_{n-2k,k}. 
\]

Set 
\[ {  \det ( X- \lambda \, { \rm Id})_{ (k)}  \over
E_{0k}} = \lambda^{ n-2k} + I_{1k} \lambda ^ {n-2k-1} + \dots +
I_{n-2k,k} . 
\] 
The functions $I_{rk}$, $r=1, \dots, n-2k$, are
constants of motion for the FKT lattice.

\begin{example} 
We consider in detail the $gl(3,{\mathbb C})$
case:
\[ 
X= \begin{pmatrix} f_1&1&0\cr g_1&f_2&1 \cr h_1&g_2 &f_3
\end{pmatrix}.
\] Taking $H_2={1 \over 2} \,{\rm tr}\, X^2$ as the Hamiltonian, and
the above Poisson bracket 
\[ 
\dot x=\{H_2, x\} 
\] gives the
following equations: 
\begin{gather*}
\dot f_1 = -g_1,\\
\dot f_2 = g_1-g_2,\\
\dot f_3 = g_2,\\
\dot g_1 = g_1(f_{1}-f_2) -h_1,\\
\dot g_2 = g_2(f_{2}-f_3) +h_1,\\
\dot h_1 = h_1 (f_{1}-f_3)
\end{gather*}
Note that $H_1=f_1+f_2+f_3$ while $H_2= {1 \over 2}
(f_1^2+f_2^2+f_3^2)+g_1+g_2$. These equations can be written in
Lax pair form, $\dot X=[B, X]$,  by taking
\[ 
B= \begin{pmatrix}   0&0&0 \cr g_1&0&0\cr h_1&g_2 &0
\end{pmatrix}.
\]

The chopped matrix is given by 
\[
\begin{pmatrix}
g_1& f_2-\lambda \cr h_1 & g_2
\end{pmatrix}.
\]

The determinant of this matrix is $h_1 \lambda +g_1 g_2-h_1 f_2$
and we obtain the rational integral \begin{gather} I_{11}={ g_1 g_2 -h_1 f_2
\over h_1} \ . \label{a8}  \end{gather}

Note that the phase space is six dimensional, we have two Casimirs
$(H_1, I_{11})$ and the functions $(H_2, H_3)$ are enough to
ensure integrability.
\end{example}

\begin{example} 
In the case of $gl(4, {\mathbb C})$
the additional integral is
\[ 
I_{21}= { g_1 g_2 g_3 - g_1 f_3 h_2 - f_2 g_3 h_1 + h_1 h_2
\over k_1} + f_2 f_3 -g_2 . 
\]
and 
\[ 
I_{11}={ g_1 h_2 + g_3 h_1  \over k_1}-f_2 -f_3
\]
 is a Casimir.

In this example the phase space is ten dimensional, we have two
Casimirs $(H_1, I_{11})$ and the functions $(H_2, H_3, H_4, I_{21})$ 
are independent and pairwise in involution.
\end{example}

\section{Bi-Hamiltonian structure}
We recall the definition and basic properties of master symmetries
following Fuchssteiner~\cite{fuch}. Consider a differential
equation on a manifold $M$  defined by a vector field $\chi$.
 We are mostly interested in the case where $\chi$ is a Hamiltonian vector
field. A~vector field $Z$ is a   symmetry of the equation  if
\begin{equation*}
[Z, \chi]=0  .
\end{equation*}
A vector field $Z$ is called a master symmetry if
\begin{equation*}
[[Z, \chi], \chi]=0 ,
\end{equation*}
but
\begin{equation*}
[Z, \chi] \not= 0  .
\end{equation*}

\noindent Master  symmetries were first introduced by Fokas and
Fuchssteiner in \cite{fokas}  in connection with the Benjamin--Ono
Equation.

A bi-Hamiltonian system is defined by specifying two Hamiltonian
functions $H_1$, $H_2$ and two Poisson tensors $\pi_1$ and
$\pi_2$, that give rise to the same Hamiltonian equations. Namely,
$\pi_1 \nabla H_2=\pi_2 \nabla H_1$. The notion of bi-Hamiltonian system  was introduced 
in~\cite{magri} in 1978.

 Another idea that will be useful  is
the Gelfand--Zakharevich scheme for a pencil of Poisson tensors.
Consider a bi-Hamiltonian system given by two compatible Poisson
tensors $\pi_1$, $\pi_2$. Compatible means that the pencil 
\[ 
\{\ ,\ \}_{\lambda}=\{\ , \ \}_{\pi_1} + \lambda \{ \ , \ \}_{\pi_2}
\] 
is Poisson for each value of the real parameter $\lambda$.
Gelfand and Zakharevich in \cite{gelfand} consider the special
case in which the pencil possesses only one Casimir. Under some
mild conditions they prove the following: 

Let 
\[
F_{\lambda}=F_0+\lambda F_1+ \dots + \lambda^n F_n .
\]
Then $F_0$ is a Casimir for $\pi_1$, $F_n$ is a  Casimir for
$\pi_2$ and, in addition, the functions $F_0, F_1, \dots, F_n$ are
in involution with respect to both brackets $\pi_1$ and $\pi_2$.

We now return to the FKT lattice. We want to define a second
bracket $\pi_2$ so that $H_1$ is the Hamiltonian and
\[ 
\pi_2 \, \nabla  H_1 = \pi_1 \,  \nabla  H_2  . 
\]
i.e.\ we want to construct a bi-Hamiltonian pair. We will
achieve  this by finding a master symmetry $X_1$ so that 
\[ 
X_1({\rm Tr}\, X^i) = i \, {\rm Tr}\, X^{i+1}. 
\]

We construct $X_1$ by considering the equation
\begin{gather} 
\dot{X}=[ Y,X]+ X^2 . 
\label{a14} 
\end{gather}

We choose $Y$ in such a way  that the equation is consistent. One
solution is 
\[ 
Y= \sum_{i=1}^n \alpha_i E_{ii} +\sum_{i=1}^{n-1}
\beta_i E_{i,i+1}, \]
where  $\beta_i=i$, $\alpha_i= i f_i + \sum\limits_{k=1}^{i-1} f_k$.

The vector field $X_1$ is defined by the right hand side of~(\ref{a14}).

For example, in $gl(4, {\mathbb C})$  the components of $X_1$ are:
\begin{gather*}
X_1(f_1)= 2 g_1+f_1^2,\\
X_1 (f_2)=3 g_2+f_2^2,\\
X_1(f_3)=  -g_2 +4 g_3 +f_3^2,\\
X_1 (f_4)=  -2 g_3+f_4^2,\\
X_1 (g_1)=  3 h_1 + g_1 f_1 + 3 g_1 f_2,\\
X_1(g_2)=  4 h_2+ 4 g_2 f_3,\\
X_1(g_3)= -h_2 -g_3 f_3 + 5 g_3 f_4,\\
X_1 (h_1)=g_1 g_2+ 4 k_1 +h_1 f_1 +h_1 f_2 +4 h_1f_3,\\
X_1 (h_2)= g_2 g_3+h_2 f_3 +5 h_2 f_4,\\
X_1 (k_1)= g_3 h_1 +g_1 h_2+k_1 f_1 +k_1 f_2 +k_1 f_3+5 k_1 f_4.
\end{gather*}

The second bracket $\pi_2$ is defined by taking the Lie derivative
of $\pi_1$ in the direction of $X_1$.
 The bracket $\pi_2$ is at most quadratic, i.e.\ in the case $n=4$
\begin{gather*}
\{ g_i, g_{i+1} \}= g_i g_{i+1}+h_i f_{i+1},\\
 \{ g_i, h_i \}= g_i h_i,\\
\{ g_{i+1}, h_i \} = -g_{i+1} h_i,\\
\{ g_i, f_i \}=-g_i f_i,\\
\{ g_i, f_{i+1} \}= g_i f_{i+1},\\
\{ g_i, f_{i+2} \} =h_i,\\
\{ g_{i+1}, f_i \}=-h_i ,\\
\{ h_i, f_i \} =-h_i f_i ,\\
\{ h_i, f_{i+2} \} = h_i f_{i+2} ,\\
\{ f_i, f_{i+1}  \}= g_i.
\end{gather*}

 This bracket was first
obtained in \cite{damianou2} using the method described above
(i.e.\ master symmetries). A closed expression for this bracket was
obtained later by Faybusovich and Gekhman in~\cite{fay}. It was
obtained using $R$-matrices and the expression takes the following
simple  form:
\begin{gather}\{ x_{ij} , x_{kl} \}=\left\{
\begin{array}{ll}
{\rm sign } (k-i) x_{ij} x_{kj} & {\rm if} \ \ j=l, \\
x_{ij} x_{il}  & { \rm if} \ \   k=i, \\
x_{ij} x_{kl} +x_{il} x_{kj} & {\rm if } \  \  i<k\le j,   \\
x_{il}   \  & {\rm if } \ \  k=j+1  . \label{a6}
\end{array}
\right. \end{gather}

  As we will
 see in the next section, there is a linear and a quadratic
 bracket on $gl(n, {\mathbb R})$ whose restriction to the FKT lattice
 coincides with the brackets $\pi_1$, $\pi_2$.

\section{A new approach}

We already mentioned in the introduction that  our starting point
is the work of A.~Meucci~\cite{meucci1,meucci2}. His work
is a discrete analogue of a procedure that produces the KdV,
Boussineq and Gelfand--Dickey hierarcies~\cite{falqui}. For
example, the KdV is bi--Hamiltonian and it can be obtained by
reducing the space of $C^{\infty}$ maps from  $S^1$ to  $gl(2,
{\mathbb R}) $.  If one considers the space of maps from $S^1$  to $
gl(3, {\mathbb R}) $ the Boussineq hierarchy is obtained. In
\cite{meucci1,meucci2} the discrete version of the
procedure is considered. The circle, $S^{1}$,  is replaced by the
cyclic group ${\mathbb Z}_d$ and one considers maps from ${\mathbb Z}_d$
to $GL(N, {\mathbb R})$. One  obtains, after reduction, equations that
are bi-Hamiltonian. In the case $N=2$ the resulting system is the
periodic Toda lattice and for $N=3$ a system studied by
Kupershmidt in~\cite{kup}. We will not get into the details of the
procedure  but rather we will use the results as our starting
point for our own  purposes. We will content with a short outline
of the constructions of \cite{meucci1} and \cite{meucci2}.  The
basic object  is the space of maps from
 ${\mathbb Z}_d$ to $GL(N, {\mathbb R})$. This  space is endowed with a structure consisting of a
Poisson manifold together with a pair of Lie-algebroids suitably
related. The next step is a Marsden--Ratiu type of reduction and
the result is a bi-Hamiltonian system with a pair of  Poisson
structures on $gl(N, {\mathbb R})$. The Lax pair of the resulting
system contains two spectral parameters and the theory of
Gelfand--Zakarevich applies. The system turns out to be integrable
with the required number of integrals. We give explicit formulas
that we have computed from \cite{meucci1}  in the case $N=3$ both
for the pair of Poisson brackets, the Lax pair and the integrals
of motion. In the case $N=4$ we  display the Lax pair and the
polynomial integrals of motion. To obtain the FKT lattice one has
to perform a further reduction to the phase space of the FKT
lattice and to obtain the rational integrals we propose a new
algorithm which is the central new result of our paper. At the
present we do not have a~Lie algebraic  interpretation of this
algorithm. We illustrate with two examples:

\begin{example}[The $\boldsymbol{gl(3, {\mathbb C})}$ case]
The phase space of the system obtained by the procedure of Meucci
is nine dimensional, i.e.\ matrices of the form
 \[ 
  \begin{pmatrix} f_1&h_2& g_3\cr g_1&f_2&h_3 \cr h_1&g_2
&f_3
\end{pmatrix} . 
\]

 We compute the pair of Poisson brackets on the extended space with variables 
 \[ 
 \{f_1, f_2, f_3, g_1, g_2, g_3, h_1, h_2, h_3 \} . 
\] 

The  Lie--Poisson bracket is defined by the following structure
matrix
\[ 
\begin{pmatrix} 0&0&0&-g_1&0&g_3&-h_1&h_2&0 \cr 0&0&0&
g_1&-g_2&0&0&-h_2&h_3 \cr 0&0&0&0&g_2&-g_3&h_1&0&-h_3 \cr
g_1&-g_1&0&0&-h_1&h_3&0&0&0 \cr 0&g_2&-g_2&h_1&0&-h_2&0&0&0\cr
-g_3&0&g_3&-h_3&h_2&0&0&0&0\cr h_1&0&-h_1&0&0&0&0&0&0\cr
-h_2&h_2&0&0&0&0&0&0&0\cr 0&-h_3&h_3&0&0&0&0&0&0 \end{pmatrix} ,
\]
and the  quadratic Poisson bracket is defined by $A- A^t$ where
$A$ is the matrix
\[ 
\begin{pmatrix} 0& g_1& -g_3& -g_1f_1& h_1-h_2& g_3f_1&
-h_1f_1& h_2f_1& 0 \cr 0& 0& g_2& g_1f_2& -g_2f_2& h_2-h_3& 0&
-h_2f_2& h_3f_2 \cr 0& 0& 0& h_3-h_1& g_2f_3& -g_3f_3& h_1f_3& 0&
-h_3f_3\cr 0& 0& 0& 0& -h_1f_2-g_1g_2& h_3f_1+g_1g_3& -g_1h_1& 0&
g_1h_3 \cr 0& 0& 0& 0& 0& -h_2f_3-g_2g_3&
 g_2h_1& -g_2h_2& 0 \cr
0& 0& 0& 0& 0& 0& 0& g_3h_2& -g_3h_3 \cr 0& 0& 0& 0& 0& 0& 0& 0& 0
\cr 0& 0& 0& 0& 0& 0& 0& 0& 0 \cr 0& 0& 0& 0& 0& 0& 0& 0& 0
\end{pmatrix} .
\]

 The Lax matrix with two spectral parameters  is given by
 \[ L_{\lambda, \mu}=
\begin{pmatrix}
(f_1+\lambda)\mu^2 &h_2-\mu^3&\mu g_3 \cr \mu g_1&(f_2
+\lambda)\mu^2&h_3-\mu^3 \cr h_1-\mu^3&\mu g_2 &(f_3+\lambda)\mu^2
\end{pmatrix}.
\]

Let $p(\lambda, \mu)={\rm det}\, L_{\mu, \lambda}$. Write
\[ 
p(\lambda, \mu) =-\mu^9+c_2(\lambda) \mu^6 +c_1(\lambda) \mu^3
+c_0(\lambda) . 
\]
Then
\[
c_2(\lambda)=\lambda^3+k_2 \lambda^2+k_1 \lambda +k_0,
\]
where
\begin{gather*}
k_2= f_1+f_2+f_3, \\
k_1= f_1 f_2+f_1 f_3+f_2 f_3+g_1+g_2+g_3, \\
k_0= f_1 f_2 f_3+f_1 g_2+f_2 g_3+f_3 g_1+h_1+h_2+h_3
\end{gather*}
and
\[
c_1(\lambda)=l_1 \lambda +l_0,
\]
where
\begin{gather*}
l_1= -h_3g_2-g_3h_1-h_2g_1, \\
l_0=-f_1g_2h_3+g_1g_3g_2-g_1h_2f_3-h_1g_3f_2-h_2h_3-h_3h_1-h_1h_2 .
\end{gather*}
Finally, 
\[ c_0(\lambda) =h_1 h_2 h_3 . 
\]

The functions $k_i$, $l_i$, $c_0$ are all in involution in the
Lie--Poisson  bracket. The functions $k_2$, $l_1$, $c_0$ are all
Casimirs.

In the quadratic bracket $k_i$, $l_i$, and $c_0$ are all in
involution. $k_0$, $l_0$ and $c_0$ are Casimirs.

Let 
\[ 
l={l_0 \over l_1}=
{-f_1g_2h_3+g_1g_3g_2-g_1h_2f_3-h_1g_3f_2-h_2h_3-h_3h_1-h_1h_2
\over -h_3g_2-g_3h_1-h_2g_1 } . 
\] Setting $h_2=h_3=0$ in $l$
we obtain $I_{11}$~(\ref{a8}).
\end{example}

\begin{example}[The $\boldsymbol{gl(4,{\mathbb C} )}$ case]
In order to give a complete comparison of the previous and the
present method of obtaining the rational invariants we consider
first  integrability using chopping.

 We consider the matrix $L$ given by
\[ L= \begin{pmatrix} f_1 &  -1 & 0 & 0 \cr
                   g_1 & f_2 & -1 &  0 \cr
                   h_1 & g_2 & f_3 & -1   \cr
                   k_1 & h_2 & g_3 & f_4   \cr \end{pmatrix}.
\]

Note that we are using $-\epsilon$ instead of $\epsilon$ in order
to get the integrals to match exactly.

In this case the chopped matrix has the form
\[ 
Ch_1 (\lambda)=\begin{pmatrix} g_1 & f_2-\lambda & -1 \cr
          h_1 & g_2 & f_3 -\lambda \cr
           k_1 & h_2 & g_3 \end{pmatrix} .
\]

The characteristic polynomial has the form 
\[ k_1 \lambda^2 +
(g_1 h_2 + h_1 g_3-k_1 f_2 -k_1 f_3) \lambda  +g_1 g_3 g_2-g_1 h_2
f_3 -h_1f_2g_3-h_1h_2+k_1f_2f_3 +k_1g_2 . 
\] We obtain the
following two rational invariants
\[ 
i_{11}={h_2 g_1 + g_3 h_1 \over k1} -(f_2+f_3) , 
\] and
\[
i_{21}={g_1 g_3 g_2 -g_1 h_2 f_3 -h_1 f_2 g_3-h_1 h_2 \over k_1} +f_2 f_3 +g_2.
\]
\end{example}

We now turn to the gentle approach, i.e.\  integrability without
chopping.
Consider the
following Lax pair with two spectral parameters:
\begin{gather}
L_{\lambda, \mu}= \begin{pmatrix} (f_1+\lambda)\mu^3 &  k_2-\mu^4
& h_3 \mu & g_4 \mu^2 \cr
                   g_1 \mu^2 & (f_2 +\lambda)\mu^3& k_3 -\mu^4 &  h_4 \mu \cr
                   h_1 \mu & g_2 \mu^2 & (f_3+\lambda)\mu^3 & k_4-\mu^4   \cr
                   k_1-\mu^4 & h_2 \mu & g_3 \mu^2 & (f_4 +\lambda)\mu^3  \cr \end{pmatrix}.
\end{gather}
Taking determinant we obtain the polynomial
\[
p_{\lambda, \mu}=-\mu^{16}+K_3(\lambda) \mu^{12}+ K_2(\lambda) \mu^8 +K_1(\lambda)\mu^4+K_0(\lambda) .
\]
We present the explicit expressions for the polynomials $K_i(\lambda)$.
\[ \bullet \quad K_3(\lambda)=K_{33}
\lambda^3+K_{32}\lambda^2+K_{31}\lambda + K_{30}, 
\] 
where 
\begin{gather*}
K_{33}=f_1+f_2+f_3+f_4 ,\\
K_{32}=g_1+g_2+g_3+g_4+ f_1 f_2+f_1 f_3+f_1 f_4+f_2 f_3+f_2 f_4+f_3 f_4 ,\\
K_{31}= h_1+h_2+h_3+h_4+f_1 f_2 f_3+f_1 f_2 f_4+f_1 f_3 f_4+f_2 f_3 f_4 +f_2 g_4 \\
\phantom{K_{31}= }{}+f_3 g_4+f_3 g_1+f_4 g_1  +f_4 g_2+f_2 g_3+f_1 g_2 +f_1 g_3 ,
\\
K_{30}=k_1+k_2+k_3+k_4 +f_1 f_2 f_3 f_4 +g_1 g_3+ g_2 g_4+f_2 h_3+f_4 h_1 \\
\phantom{K_{30}=}{}+f_1 h_2+f_3 h_4+f_2 f_3 g_4+f_3 f_4 g_1+ f_1 f_4 g_2+f_1 f_2 g_3 .
\\[3mm]
\bullet \quad K_2(\lambda)=K_{22} \lambda^2+K_{21}\lambda+K_{20} ,
\end{gather*}
where
\begin{gather*}
K_{22}=-h_2h_4-g_2k_3-g_3k_4-h_1h_3-g_1k_2-k_1g_4 ,
\\
K_{21}=g_2h_4g_3-g_1k_2f_4-f_1h_2h_4-k_1g_4f_3-k_1f_2g_4-f_1g_2k_3-h_2h_4
 f_3  -g_1k_2f_3  \\
\phantom{K_{21}=}{}+g_1g_2h_3+h_1g_4g_3+g_1h_2g_4-f_1g_3k_4-h_1f_2h_3-g_2k_3f_4 
-f_2g_3k_4-h_2k_4\\
\phantom{K_{21}=}{}- h_2k_3-h_1h_3f_4-k_2h_4-h_3k_4-h_1k_2-h_1k_3 -k_1h_3-k_1h_4 ,
\\
K_{20}= - k_2 k_4 - k_2 k_3- k_1 k_2  - k_1 k_3 - k_1 k_4  -k_3
k_4 - k_1 f_2 g_4 f_3  
 - f_1 h_2 h_4 f_3 \\
 \phantom{K_{20}=}{}+ f_1 g_2 h_4 g_3- h_1 k_2 f_4  - f_1 g_2 k_3 f_4 - f_1 f_2 g_3 k_4 - g_1 g_2 g_4 g_3
 + g_1 h_2 g_4 f_3 \\
\phantom{K_{20}=}{}- h_1 f_2 h_3 f_4 + g_1 g_2 h_3 f_4 + h_1 f_2 g_4 g_3 - g_1 k_2 f_3 f_4
 - f_1 h_2 k_3  - g_1 k_2 g_3  - f_1 h_2 k_4 \\
\phantom{K_{20}=}{} - k_1 g_2 g_4 + h_1 h_2 g_4 - k_2 h_4 f_3  - f_2 h_3 k_4
 + g_2 h_3 h_4 - g_2 g_4 k_3 - g_1 g_3 k_4 + g_1 h_2 h_3 \\
\phantom{K_{20}=}{} - h_1 k_3 f_4 + h_1 h_4 g_3  - k_1 f_2 h_3 - k_1 h_4 f_3 .
\\[3mm]
\bullet \quad K_1(\lambda)= K_{11} \lambda+K_{10},
\end{gather*}
where
\begin{gather*}
K_{11}=h_1k_2k_3+h_2k_3k_4+k_1k_2h_4+k_1h_3k_4 ,\\
K_{10}=h_1
k_2k_3f_4+f_1h_2k_3k_4-k_1g_2h_3h_4+k_1g_2g_4k_3+k_1k_2h_4f_3
+k_1f_2h_3k_4\\
\phantom{K_{10}=}{}-h_1h_2g_4k_3 +h_1h_2h_3h_4-h_1k_2h_4g_3-g_1h_2h_3k_4
+g_1k_2g_3k_4+k_1k_2k_4\\
\phantom{K_{10}=}{}+k_1k_2k_3+k_2k_3k_4+k_1k_3k_4.\\[3mm]
\bullet \quad K_0(\lambda)=-k_1 k_2 k_3 k_4. 
\end{gather*}

\begin{remark} We note that $K_3(\lambda)$ gives  the
polynomial invariants. Clearly $H_1= K_{33}={\rm tr}\,  L$.
We also have $H_2={ 1 \over 2}\, {\rm tr}\, L^2={ 1 \over 2} K_{33}^2 - K_{32}$ and
$H_3={ 1\over 3}\, {\rm tr}\, L^3={1\over 3}K_{33}^3-K_{33} K_{32}+ K_{31} $.

Finally,
$H_4={1 \over 4} K_{33}^4+K_{33}K_{31}+{ 1\over 2} K_{32}-K_{30}-K_{33}K_{32}$.

These  last relations hold provided that $k_2=k_3=k_4=0$,
$h_3=h_4=0$ and $g_4=0$.
\end{remark}

\begin{remark} The next coefficient $K_2(\lambda)$ gives
the rational invariants.
We form the quotient ${K_{21} \over K_{22}}$ and set $k_2=k_3=k_4=0$ and $h_3=h_4=0$. We obtain
\[
{ K_{21}\over K_{22}}={ -g_4 (k_1 f_3+k_1 f_2 -h_1 g_3 -g_1 h_2) 
\over -k_1 g_4} =f_2+f_3-{ (h_1 g_3+g_1 h_2) \over k_1}.
\]
This is precisely $-i_{11}$.

Similarly, we form ${ K_{20} \over K_{22}}$ to obtain
precisely $i_{21}$.
\end{remark}

\begin{remark}
The last two terms, namely
$K_1 (\lambda)$ and $K_0 (\lambda)$ become identically zero once
we set $k_2=k_3=k_4=0$, $h_3=h_4=0$.
\end{remark}

In the general case, the polynomial $p_{\lambda, \mu}$ involves
polynomials $A(\lambda)$, $B_1(\lambda), \dots, B_k(\lambda)$ and
$C_1(\lambda), \dots, C_s(\lambda)$ with $k= \left[ { (n-1) \over 2}\right]$
and $s=n-k-1$.  The polynomial $A (\lambda)$ can be used to obtain
the polynomial integrals. The polynomials $B_i(\lambda)$ give the
rational integrals using the procedure described above and the
$C_j(\lambda)$ vanish identically once we restrict to the phase
space of the FKT lattice. A detailed proof of the general case
will be given in a future publication.

\LastPageEnding

\end{document}